\documentclass[a4paper,11pt,oneside]{book}

\usepackage[latin1]{inputenc}
\usepackage{latexsym}
\usepackage[T1]{fontenc}
\usepackage{cite}
\usepackage{url}
\usepackage{tikz}
\usepackage{fix-cm}

\usepackage{graphicx}
\usepackage{epsfig}
\usepackage{amsbsy}
\usepackage{amssymb}
\usepackage{amsmath}
\usepackage{alltt}
\usepackage{float}
\usepackage{color}
\usepackage{rotating}
\usepackage{relsize}
\usepackage{calc}
\usepackage{hhline}
\usepackage{ifthen}
\usepackage{xspace}

\newcommand{\ie}{\emph{i.e.}\xspace}
\newcommand{\eg}{\emph{e.g.}\xspace}

\newcommand{\cf}{\emph{cf.}\xspace}
\newcommand{\viz}{\emph{viz.}\xspace}

\newcommand{\wrt}{with respect to\xspace}
\newcommand{\etal}{\emph{et~al.}\xspace}

\newcommand{\true}[1]{{\smaller\textsc{TRUE}\xspace}{#1}}
\newcommand{\false}[1]{{\smaller\textsc{FALSE}\xspace}{#1}}
\newcommand{\xor}[1]{{\smaller\textsc{XOR}\xspace}{#1}}
\newcommand{\andg}[1]{{\smaller\textsc{AND}\xspace}{#1}}

\usepackage{remreset}
\makeatletter\@removefromreset{footnote}{chapter}\makeatother

\newcommand{\paperref}[1]{~\textbf{(Paper \ref{#1})}}
\usepackage{xcolor,calc}

\begin{document}
\title{Design of Reversible Computing Systems\\{\Large Logic, Languages, and Circuits}}

\author{Michael Kirkedal Thomsen\\[1em]
DIKU, Department of Computer Science, \\University of Copenhagen, Denmark }

\date{July 7, 2012\\[26em]This PhD dissertation has been approved on Sep 12 2012 at the PhD School of Science, Faculty of Science, University of Copenhagen, Denmark}

% This PhD dissertation investigates garbage-free reversible computing systems from abstract design to physical gate-level implementation.
% Designed in reversible logic, we propose a ripple-block carry adder and work towards a reversible circuit for general multiplication.
% At a higher-level, abstract designs are proposed for reversible systems, such as a small von Neumann architecture that can execute programs written in a simple reversible two-address instruction set, a novel reversible arithmetic logic unit, and a linear cosine transform.
% To aid the design of reversible logic circuits we have designed two reversible functional hardware description languages:  a linear-typed higher-level language and a gate-level point-free combinator language.
% We suggest a garbage-free design flow, where circuits are described in the higher-level language and then translated to the combinator language, from which methods to place-and-route of CMOS gates can be applied.
% We have also made standard cell layouts of the reversible gates in complementary pass-gate CMOS logic and used these to fabricate the ALU design.

% In total, this thesis has shown that it is possible to design non-trivial reversible computing systems without garbage and that support from languages (computer aided design) can make this process easier.

\pagestyle{plain}

\maketitle

\clearpage
\pagestyle{plain}
\pagenumbering{roman}

\chapter*{Abstract}
Reversible computing spans computational models that are both forward and backward deterministic. These models have applications in program inversion and bidirectional computing, and are also interesting as a study of theoretical properties. The motivation for reversible computing comes, however, often from the fact that these models are information preserving.
Landauer's principle links information theory and thermodynamics; all information has some physical representation, so a loss of information must cause a thermodynamical entropy decrease, which then leads to heat dissipation to obey the law thermodynamics. A reversible computation does, thus, not have to use energy, though this is impossible to avoid in practice, due to the way computers are build.

It is, however, not always obvious how to implement reversible computing systems. The restriction to avoid information loss, imposes new design criteria that need to be incorporated into the design; criteria that do not follow directly from conventional models. 
The result is that, today, many implementations is simple reversible embeddings of conventional solutions. This is not a desirable approach, because these simple embeddings always generate garbage, which then leads to erasure of information.

In this thesis I investigate garbage-free reversible computing systems from abstract design to physical gate-level implementation. Arithmetic operations are a basis for many computing systems, so a proposed the design of a ripple-block carry adder and work towards a reversible circuit for general multiplication are important new circuits. Such arithmetic circuits have then been used in the design of two larger reversible computing systems. 
The first is a small von Neumann architecture, called Bob, that can execute programs written in a simple reversible two-address instruction set. A central part of the architecture is a novel design of a reversible arithmetic logic unit. 
The second system is an implementation of the linear cosine transform used in the H.264/ACV encoding standard.

Designing reversible logic circuits on paper can become very complex, so I have designed two reversible functional hardware description languages that can simplify the implementation process. One language is a linear-typed higher-level language, while the other is a gate-level point-free combinator language. 
These two languages can be used in a garbage-free design flow, where circuits are described in the higher-level language and then translated to the combinator language. From the gate-level language, methods of place-and-route of CMOS gates can be applied. To facilitate this last step, I have also made standard cell layouts of the reversible gates in complementary pass-gate CMOS logic and, as a test, these cells have been used to fabricate the ALU design.

In total, this thesis has shown that it is possible to design non-trivial reversible computing systems without garbage and that support from languages (computer aided design) can make this process easier. However, there is often still a need to rethink both the problem and the solution to accommodate the no-garbage approach.

\clearpage

\thispagestyle{plain}
\chapter*{Danish Resum\'e}
\hyphenation{ind-lej-rin-ger}
\hyphenation{a-rit-me-tis-ke}
\hyphenation{mo-del-ler}
Reversible beregninger d{\ae}kker over beregningsmodeller der er b{\aa}de forl{\ae}ns og bagl{\ae}ns deterministiske. Disse modeller finder anvendelse indenfor program inversion og bidirektionel beregning, men er ogs{\aa} interessante som et studie af deres teoretiske egenskaber. Motivationen bag reversible beregning kommer dog ofte fra det faktum at disse modeller er informationsbevarende. Landauers princip k{\ae}der informationsteori sammen med termodynamikken; al informations har en fysisk repr{\ae}sentation, s{\aa} tab af information m{\aa} medf{\o}re en reduktion af termodynamisk entropi, som dermed f{\o}rer til varmeafgivelse for at overholde termodynamikkens love. En reversibel beregning vil dermed ikke have et varmetab, dog er det i praksis umuligt at undg{\aa} pga. datamatens opbygning.

Det er dog ikke altid oplagt hvordan man kan implementere reversible beregningssystemer. Restriktionen om at undg{\aa} informationstab, opstiller nye designkriterier, som er n{\o}dvendigt at inddrage i designet -- kriterier som ikke f{\o}lger direkte fra konventionelle beregningsmodeller. Resultatet er at mange implementationer  i dag er simple reversible indlejringer af konventionelle l{\o}sninger. Dette er ikke en {\o}nskelig fremgangsm{\aa}de, da disse simple indlejringer altid vil generere ``affald'', som efterf{\o}lgende medf{\o}rer informationstab.

I denne afhandling unders{\o}ges affaldsfrie reversible beregningssystemer fra abstrakt design ned til implementation p{\aa} fysisk port-niveau. Aritmetiske operationer er grundlaget for mange beregningssystemer, s{\aa} der foresl{\aa}s et design til et ``ripple-block carry'' additionskredsl{\o}b og forel{\o}bigt arbejde mod et reversibelt kredsl{\o}b for general multiplikation er vigtige nye kredsl{\o}b. S{\aa}danne aritmetiske kredsl{\o}b er brugt i designet af to st{\o}rre reversible beregningssystemer. Det f{\o}rste er en lille von Neumann arkitektur, kaldet Bob (eng. for lod), som kan udf{\o}re programmer skrevet i et simpelt reversibelt to-adresse instruktionss{\ae}t. En central del af arkitekturen er et nyt design af en reversibel aritmetisk logisk enhed (ALU). Det andet system er en implementation af en line{\ae}r cosinus transformation, som bruges i H.264/AVC videokodningsstandarden.

Design af reversibel logik p{\aa} papir kan nemt blive meget komplekst, s{\aa} jeg har udviklet to reversible funktionelle hardwarebeskrivelsessprog, som kan simplificere implementationsforl{\o}bet. Det ene er et line{\ae}rt-typet h{\o}j-niveau sprog, mens det andet et `logisk port'-niveau punkt-frit kombinatorsprog.  Fra port-niveau sproget kan s{\aa} benyttes metoder til placering-og-rutning af CMOS porte. For at facilitere det sidste skridt, har vi ogs{\aa} lavet standard celle layout af de reversible porte i komplement{\ae}r pass\'er-port CMOS logik, og, som en test, har disse celler v{\ae}ret brugt til at fabrikere ALU designet.

Alt i alt har denne afhandling vist at det er muligt at designe ikke-trivielle reversible beregningssystemer uden affald, og at hj{\ae}lp fra programmeringssprog (datamat st{\o}ttet design)  kan g{\o}re dette forl{\o}b nemmere. Der er dog stadig ofte et behov for at gent{\ae}nke b{\aa}de problemet og l{\o}sningen for at akkommodere \emph{intet-affald fremgangsm{\aa}den}.

\clearpage
\pagestyle{plain}
\tableofcontents

\clearpage
\chapter*{\vspace{-25mm}Preface}

This thesis has been submitted to the PhD School of Science, Faculty of Science, University of Copenhagen in partial fulfillment of the requirements for a PhD degree at DIKU, Department of Computer Science, University of Copenhagen, Denmark.

The thesis is written as a synopsis with 11 annexed research papers. The first chapter of the synopsis contains a short introduction to reversible computation and a description of the topic and objectives of the thesis. Following this, are three chapters that detail the contributions of my research and its relation to existing knowledge. The synopsis ends with conclusions and perspectives for future research. References to related work are given throughout the synopsis. Five appendices contains the 11 research papers; of these there are (at the time of writing) 6 published journal and conference papers, 2 pre-print conference papers (accepted for publication), 2 workshop papers, and 1 technical report.

This thesis is not the result of one person's lone fight against the world and it's reckless handling of information. Therefore, I would like to thank my advisor Robert Gl\"uck, who introduced me to the subject, and my co-author, office mate, and friend Holger Bock Axelsen.
I also thank my other co-authors (the Flemish connection) Alexis De Vos, St\'ephane Burignat, Kenneth Vermeirsh, Micha\l\ Klimczak, and Mariusz Olczak.
The collaboration with Oticon A/S has been important to the direction of the research and, from there, I would especially like to thank Kenneth Branth and Kim Poulsen.
Many other people have had an influence on this thesis, so in arbitrary (alphabetic) order I would like thank Jesper Louis Andersen, Patrick Bahr, Poul Johannis Clementsen, Fritz Henglein, Mathias Horn, Susan Nasirumbi Ipsen, Mathias Lehnfeld, Lars Valdemar Mogensen, Torben \AE. Mogensen, Kenji Moriyama, Jette Giovanni M{\o}ller, Tho\-mas P\'ecseli, Claus R{\o}rbech, Mary Sheeran, Jens Spars{\o}, Christen Artagnan S{\o}rensen, Robert Wille, Tetsuo Yokoyama, and rest of APL group at DIKU.
Also, I would like to thank the Danish Strategy Research Council for funding the MicroPower research project and, thus, my PhD stipend. 

Finally, I thank my family and friends for their support during this period. Without this, I would not have been able to do all the work and travel.

~

\noindent For Augusta.

{
\flushright\normalfont{\line(1,0){150}\\Michael Kirkedal Thomsen}\\

}

\cleardoublepage

\pagenumbering{arabic}
\setcounter{page}{1}

%%%%%%%%%%%%%%%%%%%%%%%%%%%%%%%%%%%%%%%%%%%%%%%%%%%%%%%%%%%%%%%%%%%%%%%%%%%%%%%%
%%% INTRODUCTION
%%%%%%%%%%%%%%%%%%%%%%%%%%%%%%%%%%%%%%%%%%%%%%%%%%%%%%%%%%%%%%%%%%%%%%%%%%%%%%%%

\chapter{Introduction}
\label{sec:Introduction}
\emph{Reversible computing} was introduced
by Bennett~\cite{Bennett:1973} and concerns (universal) computation models where a result can not only be computed, but also uncomputed. We also define these as models that are both forward and backward deterministic.
Though reversible computation models can compute all \emph{injective computable functions}, injectivity is not enough to characterize a reversible computing model; we must also require that each computation step is bijective.

This important requirement provides the connection to preservation of information, which is a key motivation for research in reversible computing.
A motivation that has its foundation in 1961 with a principle defined by Landauer~\cite{Landauer:1961}; a principle that was experimentally verified very recently~\cite{BerutEtal:2012:Nat}.

\section{Information and the Limit of Computation}
\label{sec:Int:Landauer}
The search to understand the computation process and its limitations is older that computers themselves.
Here, we do not think of the algorithmic bounds (which is a very interesting topic by itself), but the limitations that are imposed by the physical world. All computers are situated in the physical world, so the laws of physics, thus, also apply for computers including their circuits and memory. The question is what impact do the laws of physics have on the computation process?

During the 1950's (only shortly after the invention of the modern electronic digital computer~\cite{BurksGoldstineNeumann:1947}) the assumption arose that a logic operation requires an energy dissipation of $kT\ln 2$, where $k$ is Boltzmann's constant and $T$ is the temperature at which the operation is performed\footnote{This assumption followed from earlier work by Szilard~\cite{Szilard:1929} and Shannon~\cite{Shannon:1948} that argued that communicating one bit required this dissipation (\cf~\cite{Landauer:1994}).}. Von Neumann has later been credited for saying that this amount of energy is dissipated by two different sources: namely ``per elementary decision of a two-way alternative and per elementary transmittal of one unit of information''~\cite{Neumann:1966:TSA}\footnote{John von Neumann died in 1957; four years before Landauer's seminal paper. The paper was finished by Burks and published in 1966, but the work still dates back to the 1950's.}. From the first source, it \emph{is} apparent that von Neumann meant something less than all computations and it sounds a lot like a conditional, which we today know to be a problem. Today, we also know that the second source does not necessitate energy dissipation (\cf~\cite{Landauer:1991}).

The breakthrough came in 1961 when Landauer~\cite{Landauer:1961} argued that all logical operations are associated with a physical operation and because physical irreversibility requires heat generation, then so does logical irreversibility. It is, thus, not the computation process itself that necessitates energy dissipation, but the process of deleting information. Today, we refer to this as \emph{Landauer's principle}, and the dissipation of $kT\ln 2$ Joules per deleted bit of information is called \emph{Landauer's limit}\footnote{In practice the dissipation per bit is proportional to the signal energy used to represent the bit~\cite{Gershenfeld:1996} and the actual dissipation is, thus, expected to be higher.}.
Landauer founded his principle on a thought experiment in which each bit of information is encoded in a single particle. This is extremely hard to implement in the physical world, nonetheless, Landauer's principle was experimentally verified very recently~\cite{BerutEtal:2012:Nat}.

At the time of Landauer's paper (and in the following decade) it was, however, believed that erasing information was an unavoidable consequence of a computation process. Landauer had realized that all irreversible operations can be embedded in a reversible operation and, as an example, he embedded the \textsf{AND} gate in the reversible gate that we today call a \emph{Toffoli gate}. He, however, only imagined that these \emph{Landauer embeddings} could be used to temporally store the inputs of each gate, which then had to be deleted later, thus, leading to irreversibility.

\section{Foundations of Reversible Computing}
\label{sec:Int:RCfound}
The second breakthrough came in 1973 with Bennett's seminal paper~\cite{Bennett:1973}, where he defined the first universal reversible computation model; he constrained the conventional (irreversible) Turing machines (TMs) to define the \emph{reversible Turing machines} (RTMs)\footnote{The first mention of reversible Turing machines can be dated back to Lecerf in 1963\cite{Lecerf:1963}.}. In this paper Bennett also demonstrates how to embed an irreversible TM in an RTM using a history tape (similar to Landauer's embedding) and then run this RTM with a compute-copy-uncompute method (today we call this \emph{Bennett's method}) such that the overall result is only the input and the calculated output. This is a significant improvement over the use of a general trace, but often we are only interested in the result of a calculation and not both the input and output.
Bennett later showed~\cite{Bennett:1989} how, for injective functions, this input can be uncomputed by using more time, \viz adding an extra compute-copy-uncompute phase.
This is a very important result. In our research, we seek here to completely avoid garbage, because we need to know what is possible within the computational models. If needed, for a practical implementation perspective, it is easier to relax this criteria than it is strengthen it.
Further research in the tradeoff between time and space on one hand, and erasure of information (garbage) on the other have been performed by Bennett, Levine, Sherman, Vitanyi, and more~\cite{BennettEtal:1993:TCID,LevineSherman:1990,Vitanyi:2005,YokoyamaAxelsenGlueck:2011:MVLSC,ChauLo:1997}.

After Bennett's seminal paper, focus in reversible computation was directed towards more practical models (see Sections~\ref{sec:Int:QC_RL} and \ref{sec:Int:RevLang}). But at the beginning of the 1990's, interest in the theoretical aspects increased~\cite{Bennett:1989,LevineSherman:1990,JacopiniEtAl:1990,Pin:1992} and since then \emph{computability and complexity} of the reversible model has developed into a research area by itself; trying to find the place within the \emph{Complexity Zoo}~\cite{Aaronson:CompZoo} for the reversible complexity classes.
This has lead to research both on RTMs~\cite{LangeMcKenzieTapp:2000,AxelsenGlueck:2011:LATA,Axelsen:2012:LNCS,CrescenziPapadimitriou:1995,BuhrmanTrompVitanyi:2001:PhA,AxelsenGlueck:2011:FoSSaCS} and different models of \emph{reversible automata} \cite{KutribMalchar:2010,Axelsen:2012:LATA,Morita:2001,MoritaEtal:2005,Morita:2008,MoritaYamaguchi:2007}. An interesting result being that reversible space equals deterministic space~\cite{LangeMcKenzieTapp:2000}.

\section{Reversible Logic and Quantum Computing}
\label{sec:Int:QC_RL}
From the beginning, logic has had a central place in the ideas of reversible computing; \eg Landauer's ideas for the reversible gates were designed as a method to reduce the heat dissipation of logic circuits.

Inspired by Landauer's and Bennett's work, Fredkin and Toffoli had, in 1978, been working to design a reversible computer that should be based on \emph{conservative logic}~\cite{FredkinToffoli:1982}\footnote{Fredkin and Toffoli's paper was published in 1982, but the paper was based on internal papers and an MIT course from 1978; see \cite[References]{Toffoli:1980:lncs}}. In conservative logic all logic gates must be both reversible and parity preserving; \ie the number of \true values must also be preserved over the gate. For this they introduced the \emph{Fredkin gate}, which can be characterized as a controlled-swap gate\footnote{The gate that Fredkin and Toffoli presented swapped the two input-values if the control is \false. The Fredkin gate that we use today swaps if the control is \true.}.
The model was developed to reflect fundamental physical principles and they developed a \emph{billiard ball model} with a physical representation of the Fredkin gate.
Conservative logic is, however, a stricter model than reversible logic, so in 1980 Toffoli presented the \emph{$n$-bit controlled-not gates}~\cite{Toffoli:1980:lncs}. This is the most widely used class of reversible gates today, because the gates have a simple mathematical definition, which makes reversible logic synthesis easier (see Section~\ref{sec:CAD}). The class covers the \emph{not}, \emph{Feynman}, and \emph{Toffoli gates}. To follow the idea of the Fredkin gate, Toffoli also presented ingenious physical designs of these gates~\cite{Toffoli:1981} based on `differential gears' to implement the \xor and a `rotating cam' to implement the \andg in the Toffoli gate.

At the same time as Fredkin and Toffoli's work, other people were developing ideas for another radically new computer design. In 1980 Benioff presented his paper on how to design a (classical) computer from quantum components~\cite{Benioff:1980}, which shortly after (in 1982) was followed by Feynman's paper on a computer that can simulate quantum physics~\cite{Feynman:1982}. In 1985 Deutsch presented his \emph{universal quantum computer}~\cite{Deutsch:1985} and the new field was born.

In the quantum model, it was easy to include Fredkin and Toffoli's gates and in 1985 Feynman, with his flair for intuitive graphical descriptions, introduced the diagram notation that is used today~\cite{Feynman:1985}. In this paper, Feynman also introduced the very first reversible arithmetic circuits; these circuits include a full-adder implementation with four reversible gates. More (reversible) quantum gates (\eg~\cite{Peres:1985}) and different notation were introduced, so in 1995 Barenco, Bennett, Shor, and others worked as a `standardization committee' and decided on the notation and a set of universal quantum gates~\cite{BarencoEtAl:1995}.

From a historical perspective, the idea of reversible logic circuits are even older than Fredkin and Toffoli's work. In 1959, two years before Landauer's paper, Huffman looked at \emph{information-lossless} finite state machines (FSMs)~\cite{Huffman:1959}. He was interested in signal transformation (both for encoding and cryptography) and for these applications information-lossless FSMs are perfect; by constructing the encoding machine you also get the decoding machine. To design these FSMs, he defined information-lossless gates similar to reversible gates. Huffman also showed that adding a control signal calculated by an irreversible function to a reversible gate, does not break reversibility of the gate. The circuits are, however, not reversible according to our definition, but this was also not his purpose.

\section{Reversible Programming Languages and Transformation}
\label{sec:Int:RevLang}
Another track in the history of reversible computing begins in 1986. At this time, Lutz, after a brief meeting with Landauer, sent him a letter about some work he did with Derby, about four years earlier, on a reversible imperative language called Janus~\cite{LutzDerby:1986}. Their work arose from an interest to investigate if it was possible to implement such a language and before 1986 Lutz and Derby did not know about Landauer's principle. The language was `rediscovered' after the turn of the century and has since then been formalized and further developed at DIKU~\cite{YokoyamaGlueck:2007:Janus,YokoyamaAxelsenGlueck:2008,ClementsenEtAl:2010:NWPT}; here, also student projects that implement more advanced algorithms (\eg matrix multiplication) have been made~\cite{MadsenPoulsen:2011:StudRep,JensenJensen:2011:StudRep}.
Other simple reversible imperative languages have been developed, \eg Frank developed R~\cite{Frank:1999} generate instruction code for the Pendulum processor and Matos~\cite{Matos:2003} made a language for linear programs.

Though the first reversible programming language was imperative, reversible functional languages have lately received the most interest. This development started in 1996 when Huelsbergen presented SECD-H~\cite{Huelsbergen:1996}, an evaluator for the lambda calculus that extended Landin's SECD evaluator~\cite{Landin:1964} with a history tape. (Kluge~\cite{Kluge:2000} similarly extended a machine that can reduce a program term to normal-form and back again.) This was followed by Mu \etal who, with applications in bidirectional computing in mind, presented a reversible relational language~\cite{MuHuTakeichi:2004}. 
More recently, work towards general purpose functional programming languages was presented independently by Yokoyama, Axelsen, Gl\"uck~\cite{YokoyamaAxelsenGlueck:2012:LNCS} and James, Sabry~\cite{JamesSabry:2012:POPL}.

Also, a variety of languages for modeling quantum computations have been designed in almost all different paradigms. We will not detail these here, but refer to Gay's survey of the area~\cite{Gay:2006:QPL}.

On a related topic, transformation of reversible languages have also received some interest lately (mainly at DIKU). Though the first compiler between two reversible language was made by Frank~\cite{Frank:1999} (between his language R and a reversible instruction set called PISA), it was Axelsen who devised techniques for a compiler that could perform clean translation~\cite{Axelsen:2011:CC}. His translation was between Janus and PISA and was clean in the sense that the compiled program did not have more than a constant memory overhead over the original Janus program. It is likely that the PISA program will use more temporary memory/registers.
Mogensen's also did work on partial evaluation of Janus~\cite{Mogensen:2011,Mogensen:2012}.

We would also like to mention (automatic) program inversion. Inverting a reversible program is often easy, but if the program is implemented in a irreversible language it is harder. Program inversion has a direct impact on program maintainability and reliability of the inverse programs, which are otherwise hard to find if the program to be inverted either is formulated in a conventional programming language~\cite{MatsudaEtal:2010,SrivastavaEtal:2011}, or have to be deduced by static program analysis techniques~\cite{GlueckKawabe:2005:LR} or interpretation~\cite{AbramovGlueck:2002}. 
We should also mention semi-inversion, where the inverted program not necessarily is a mapping from output to input, but a combination of the original inputs and outputs, \cf Mogensen~\cite{Mogensen:2005,Mogensen:2008}.

Work on reversible instruction sets is covered in Section~\ref{sec:LogicDesign}, while design language for reversible logic is discussed in Section~\ref{sec:cad:designLang}.

\section{Towards Reversible Computer Organization and Design}
In his 1961 paper, Landauer wrote that ``we shall label a machine as being logical reversible, if and only if all its individual steps are logically reversible''~\cite{Landauer:1961}. This is a very grand challenge and we know from Bennett (and later work) that, theoretically, such machines \emph{do} exist -- even when we add the requirement that the final result must not include garbage.
But is it possible to realize such machines in practice and can it be done with the fabrication technology that exists today? And will we actually be able to achieve the expected reduced heat dissipation?

\begin{figure}
\includegraphics[width=\textwidth]{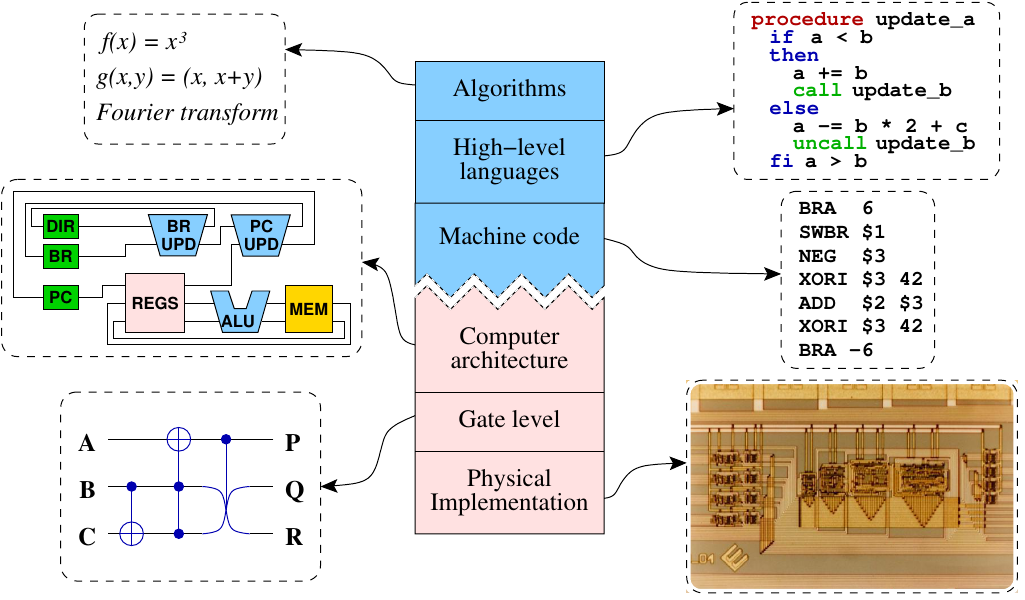}
\caption{Tower of reversible computing system~\cite{AxelsenEtAl:2009:ERCIM}.}
\label{fig:tower}
\end{figure}

The \emph{MicroPower} research project~\cite{AxelsenEtAl:2009:ERCIM}, which started in 2009, has as objective to develop a \emph{proof-of-concept reversible computing system} and the computer science theory behind it. To do this all parts of the \emph{reversible computation tower} (Figure~\ref{fig:tower}) must be investigated.
More specifically, the project investigates whether reversible computing can be applied in a power-limited application (specifically hearing aids) with the future hope to either reduce power consumption or increase functionality.

This dissertation is part of the MicroPower research project. Within this project, my thesis is that making a garbage-free reversible computing system is not only feasible, but does not necessarily require much more effort than making a conventional computing system. We will \emph{not} address the questions of the actual energy consumption of the circuits and the theoretical implications; these two fundamental questions are investigated by other parts of the project.

To answer our thesis we must investigate the bottom part of the computing tower (Figure~\ref{fig:tower}) from the machine code level down to the physical implementation. We investigate and design basic reversible logic circuits with an emphasis on arithmetic (Section~\ref{sec:logic:arithmetic}). These are important basic operations in all computing systems and will, therefore, give the foundation for the reversible systems. Improvements here will improve all other parts of the tower.
We also look at reversible circuits from a higher abstraction, namely in terms of logic designs of reversible computer architectures (Section~\ref{sec:logic:architecture}) and multimedia transforms (Section~\ref{sec:logic:transform}). These two applications have the potential to be the first practical use of reversible circuits: architectures to create very small stand-alone systems (\eg sensors) and multimedia transforms, which can be embedded in an irreversible system.

We investigate different \emph{computer aided design} (CAD) approaches for reversible circuits (Section~\ref{sec:CAD}). Logic circuits designed by hand are often efficient, but it is very time consuming and verification of the designs are not practical. The purpose of a hardware description language is to raise this abstraction.
Finally, we investigate how to implement reversible circuits in CMOS (Section~\ref{sec:circuits}). Here we desire implementations that can be used with the CAD process, but still have the prospect of reduced energy consumption. We will also look into how reversible CMOS circuits can be embedded in `conventional' static CMOS circuits.
Finally, we will look at future research topics (Section~\ref{sec:conclusion}).

%%%%%%%%%%%%%%%%%%%%%%%%%%%%%%%%%%%%%%%%%%%%%%%%%%%%%%%%%%%%%%%%%%%%%%%%%%%%%%%%
%%% 
%%%%%%%%%%%%%%%%%%%%%%%%%%%%%%%%%%%%%%%%%%%%%%%%%%%%%%%%%%%%%%%%%%%%%%%%%%%%%%%%
\chapter{Design of Reversible Computing Systems}
\label{sec:LogicDesign}
To avoid the heat dissipation from Landauer's principle the entire computing system must be fully reversible. In this chapter we will look at designs of reversible computing systems at and close to the logic gate level. We first focus on design of arithmetic logic circuits. Then, based on these, we look at reversible computing architectures and designs of multimedia transforms.

\section{Arithmetic Logic Circuits}
\label{sec:logic:arithmetic}
Arithmetic operations lie at the foundation of most computing systems and good logical implementations of these are important. Improvements to arithmetic circuits can result in improvements to the entire computing system.
In a garbage-free reversible computing system it is especially important that the arithmetic circuits are also garbage-free, but how to do this is not always obvious, and history shows that rethinking our current knowledge can be necessary.

\subsection{Addition}

An immediate problem for reversible adder implementation is that
addition is \emph{not} an injective function in itself: given just
the value of the sum $A+B$, one can not determine $A$ and $B$
uniquely. We can, however, redefine the problem by using \emph{reversible updates}~\cite{AxelsenGlueckYokoyama:2007} and to avoid overflow we use $n$-bit \emph{modular addition}, $(A,B) \mapsto (A, B + A \;mod\; 2^n)$, to define reversible addition.

The adder that Feynman proposed~\cite{Feynman:1985} was a reversible \emph{embedding} of the ripple-carry adder.
Though addition is an injective function if one of the inputs is kept, the conventional ripple-carry structure is \emph{not} reversible. The problem lies in the use of the full-adder circuit, because it is not possible to calculate both the sum and the carry without copying one of the inputs. You can say that there is an overlap in the information contained in the two results and this results in a garbage bit.
Several reversible adder designs used this ripple structure; \eg reversible \emph{binary-coded decimal} adders have received some interest~\cite{ThomsenGlueck:2008}\paperref{paper:JSA}.

The solution to this problem was presented in 1996 by Vedral \etal~\cite{VedralBarencoEkert:1996}. They observed that, to do an addition it is not necessary to calculate both the carry and the sum at the same time. If we first calculate the carry in a normal ripple, we can then make a \emph{backwards ripple} where we both uncompute the carry and calculate the sum. This was the first example that shows that it really pays to rethink the problems that we want to solve. However, the cost of not creating garbage was an increase in logic depth.
Vedral's \emph{V-shaped adder} was a huge improvement (produces no garbage), but it was not optimal \wrt ancillae and logic depth. In 2005 two papers were published that suggested different improvements to the adder design (in terms of ancilla, gate count and logic depth)~\cite{CuccaroEtAl:2005,VanRentergemDeVos:2005}.
See~\cite{ThomsenAxelsen:2009:PPL} for a detailed description of the adder circuits.

The \emph{V-shaped} adder is a ripple-carry adder and therefore has large logic depth. We know from conventional logic that it is possible to implement a logarithmic-depth adder at the cost of more logic gates and a more complicated circuit. The first reversible implementations of these adders were embeddings of the carry-lookahead adder~\cite{DesoeteDeVos:2002,KhanPerkowski:2007} and the parallel prefix adder~\cite{FeinsteinEtal:2007}, but they all suffered from garbage generation. Based on the techniques from the carry-lookahead adder, Draper \etal~\cite{DraperEtAl:2008} in 2008 presented a garbage-free logarithmic-depth adder (QCLA)\footnote{The adder was designed with a focus on quantum computing and, thus, the `Q' in the name. However, it is only implemented with reversible gates.}, which added an extra `look-behind' phase.

We presented another approach to a faster-than-linear reversible adder, the year after, named \emph{ripple-block carry-adder} (RBCA)~\cite{ThomsenAxelsen:2009:PPL}\paperref{paper:PPL}. Basically it is a ripple-carry adder but instead of calculating and rippling one bit at a time, the addition is divided into several smaller block-additions that is performed in parallel. During computation a \emph{carry-correction} phase is added and it is this phase that contains a ripple. When choosing the optimal block-size in relation to the input-size, the adder has a logic depth that is the square-root on the input-size.

When comparing the two, the QCLA in faster than the RBCA, but only significantly when the input is larger than 32 bits. In terms of gate-count the two adders are comparable, but the RBCA uses less ancilla bits. Also the RBCA has a better `locality' (the gate uses wires that, in the diagram notation, are closer to each other), which can have an impact when the circuit is implemented in the target technology (\eg quantum computer or CMOS).

\subsection{Multiplication}
Addition has an intuitive reversible formulation using modular arithmetic and reversible updates. Multiplication, on the other hand, is more difficult to define; mainly because the inverse operation is division. A simple way to define it is to take the embedding from Bennett and save both the multiplicand and the multiplier while still producing the product. This is the trick used by Kawada~\etal in their (garbage-free) reversible logic implementation of the Karatsuba algorithm~\cite{KowadaEtal:2008:RKA}.

This approach is, however, not satisfying because it expands the amount of information; on the other hand, it is not in general possible to only update the multiplicand and save the multiplier. A possibility (also suggested by Ressler~\cite{Ressler:1981:MT}) is to add a remainder, such that $mult(A,M,R) = (A*M+R,M)$, where $0 \leq R < M$. We have started the work towards a reversible logic implementation~\cite{AxelsenThomsen:2012:RC}\paperref{paper:RC2012:Mult}. So far, it works for a certain class of constant multipliers that are equal to $2^k \pm 1$ for $k \in \mathbb{N}$. We expect that this can be extended to general multiplication.

\section{Computing Architectures and Instruction Sets}
\label{sec:logic:architecture}
As early as 1965, Reilly and Federighi presented a small instructions set, designed for a one-address machine with accumulator and multiplier-quotient, with the purpose of implementing reversible subroutines~\cite{ReillyFederighi:1965}. The set is, however, not reversible \wrt our definition;
\eg it includes a clear-accumulator instruction, which does not delete information from the overall system but still dissipates energy due to the clear process. The motivation was also not energy efficiency, but a desire to enable code sharing of the implemented subroutines.

The work by Ressler\cite{Ressler:1981:MT} from MIT in 1981 was, however,  motivated by energy effficiency. Supervised by Fredkin, he made the design of a conservative logic computer based on a two-address von Neumann architecture. The instruction set contains the basic reversible arithmetic/logic instructions (\eg add, subtract, roll-left and right), and the work contained the idea of using paired-branch instructions and to have both condition and assertion in the conditional. On the other hand, memory could be irreversibly updated and (perhaps most importantly from a program inversion perspective) there was no possibility for inverse execution (uncall).

In 1987 Briggs~\cite{Briggs:1987} described a system to control an electronic cricket scoreboard. An important feature of the system was the ability to `undo' the operations that were performed; a nice feature if the operator made an error. The reverse execution was based on a trace that contained only the minimal information to do the backtracking.
Based on this idea of a minimal trace, Cezzar~\cite{Cezzar:1991} in 1991 presented (and patented) a two-address `general purpose' machine that can reverse its forward execution.

Having a trace (which would be as long as the number of executed instructions plus amount of deleted information) is, however, a very unsatisfactory solution for a reversible computer.
While Cezzar was mainly interested in a computer that could backtrack, Hall~\cite{Hall:1994} designed his improved architecture with reversibility in mind.
This ISA does not use a trace, but instead jumps are handled by come-from instructions, which makes it possible to break reversibility.

At MIT, Knight and Younis~\cite{YounisKnight:1994} had been working on a energy efficient logic family (see Section~\ref{sec:circuits}) and this implementation technology revived the interest in reversible architectures. Over a few years (ending in 1999) Vieri and Frank developed the reversible von Neumann architecture Pendulum~\cite{Frank:1999,Vieri:1999,Vieri:1995} (formalized in~\cite{AxelsenGlueckYokoyama:2007}).
This architecture was a big step forward. Instead of using only a single special-purpose register for program control (the program counter), the address calculation of the reversible abstract machine relies on \emph{three} special-purpose registers: \emph{program counter}, \emph{branch register}, and \emph{direction bit}. The calculation of the next program counter is then only dependent on the branch register and the direction bit,
and the individual instructions cannot directly alter the program counter, but instead only update the branch register and direction bit.
Though Pendulum has been fabricated, there is no full detailed description of the actual logic implementation and there is some indication that it is not fully reversible. Firstly, the implementation technology~\cite{YounisKnight:1994} (see Section~\ref{sec:circuits:adiabatac} for more details) does not implement the reversible gates, but uses Bennett's method, which results in both the input and output. Secondly, there is no use of the garbage-free V-shaped adder or other arithmetic circuits; Vedral \etal~\cite{VedralBarencoEkert:1996} is not mentioned by Frank and Vieri, and the rest of the research is from after 1999. 
Thus, although the abstract architecture of Pendulum (as described by Frank) is reversible, it is likely that the logic implementation is not.

Inspired by the Pendulum architecture and it instruction set, we have designed a fully reversible and garbage-free two-address von Neumann architecture called Bob~\cite{ThomsenAxelsenGluck:2012:LNCS}\paperref{paper:Bob}.
It features a locally-invertible instruction set, and the design, including areversible control logic and address calculation, is simple enough to be directly implemented in reversible logic.
A central part of the processor design is the arithmetic logic unit (ALU).
The conventional ALU has an inherently irreversible functionallity so we have suggested a novel alternative design for a reversible ALU~\cite{ThomsenGluckAxelsen:2010}\paperref{paper:JPhysA}.
The design of the ALU in based on the V-shaped adders and follows a strategy that puts all logical operations in sequence and then uses controls to ensure that only the desired operation changes the input. To our knowledge, this is the first garbage-free ALU.

\section{Multimedia Transformation}
\label{sec:logic:transform}
Multimedia transforms are an interesting application area for reversible circuits. In small battery-powered devices (\eg smartphones and mp3-players) they are often included as part of an ASIC to reduce power consumption and a key property of many such transforms is that they are information-lossless (and thus invertible). There exist many application areas of such transforms and even the earliest quantum algorithms (including Shor's factorization algorithm~\cite{Shor:1994}) make use of a quantum implementation of the Fourier transform. Also a implementation of the fast Fourier transform in reversible logic has been investigated~\cite{SkonecznyRentergemDeVos:2008}.

Our contributions in this area have focused on implementation of wavelet transforms in reversible logic. Wavelet transforms have also been implemented in quantum computing~\cite{FijanyWilliams:1999}, but our work builds on a paper by Bruekers and van den Enden~\cite{BruekersEnden:1992}. Here, they showed a new network structure (the so-called \emph{lifting scheme}) that can be used for perfect inversion and reconstruction of the inputs. This is desired in many transforms and the properties are also a perfect match for reversible implementations. Daubechies and Sweldens show how to factorize wavelet transforms into a lifting scheme~\cite{DaubechiesSweldens:1998} and we use this to find the lifting scheme for the linear transform of the \emph{H.264 video encoding}~\cite{DeVosBurignatThomsen:2012:MVLSC}\paperref{paper:Multimedia1}. This implementation, however, generates garbage that is caused by a multiplication-by-5.\footnote{The design was made before our own garbage-free constant multiplication circuit~\cite{AxelsenThomsen:2012:RC}, which solves the problem with multiplication-by-5.} In the latter work we examine other linear transforms that only have multiplications-by-2~\cite{BurignatEtal:2012:RC}\paperref{paper:Multimedia2} and find the associated lifting scheme. (The papers also contain work on CMOS fabrication and testing of the transform from \cite{DeVosBurignatThomsen:2012:MVLSC}. We shall discuss this in Section~\ref{sec:circuits}.)

%%%%%%%%%%%%%%%%%%%%%%%%%%%%%%%%%%%%%%%%%%%%%%%%%%%%%%%%%%%%%%%%%%%%%%%%%%%%%%%%
%%% Design Languages
%%%%%%%%%%%%%%%%%%%%%%%%%%%%%%%%%%%%%%%%%%%%%%%%%%%%%%%%%%%%%%%%%%%%%%%%%%%%%%%%
\chapter{Computer Aided Design of Reversible Circuits}
\label{sec:CAD}

In Section~\ref{sec:LogicDesign} we discussed the design of reversible logic circuits. All presented circuits (adders, multipliers, and transforms) have a very regular structure, which makes it a lot easier for humans to reason about them. Also, much of the development comes from novel design ideas based on a theoretical insight to the problem.
However, not all problems have these properties, so in order to make good realizations of more complex circuits possible, efficient description languages, logic synthesis, and optimization techniques are being developed. Computer aided design of Boolean circuits has been (and is still being) developed (\cf \cite{Micheli:1994:SODC}), so in this section we will focus on methods to aid the design of reversible circuits.

\section{Reversible Logic Synthesis and Optimization}
\label{sec:cad:synthesis}
The first approach to reversible logic synthesis is actually a very beautiful example of how mathematics can be related to reversible circuits.
Based on work by Rayner and Newman~\cite{RaynerNewman:1995}, Storme and De Vos~\cite{StormeEtAl:1999} used that reversible gates and cascading of these by serial composition forms a group, with the result that it is possible to use the known methods from group theory.
Specifically, one of these methods can be used to decompose an arbitrary reversible circuit into a cascade of simpler circuits that only updates one input wire~\cite{DeVos:2010:book,DeVosEtAl:2006}. Each of these simpler circuits can then be interpreted as an \emph{exclusive-or-sum-of-products} (ESOP), which basically is a cascade of controlled-not gates with the number of controls of each gate equals to the number of variables in the products.

Reversible logic synthesis have since been much researched and often with an interest to also apply the methods to quantum circuits. Many of the approaches is based on techniques known from Boolean logic synthesis. Perkowski \etal~\cite{PerkowskiEtal:2001,Al-Rabadi:2004} made a method for hierarchical decomposition using three known decomposition techniques and
Shende \etal~\cite{ShendeEtal:2002:RLCS} implemented a brute-force algorithm with memorization, that can find optimal circuits. Maslov, Dueck, and Miller~\cite{MaslovDueckMiller:2003,MaslovDueckMiller:2005} implemented heuristic methods based on truth tables that also use the Fredkin gate. Other approaches are based on positive-polarity Reed-Muller expansion~\cite{GuptaEtAl:2006}, or Reed-Muller spectra~\cite{MaslovDueckMiller:2007}, and finally optimal synthesis based on satisfiability of Boolean formulas~\cite{WilleEtAl:2008,GrosseEtal:2007}. There exists many more, but common to these are that they are based on truth tables as input and that they actually try to solve an NP-hard problem. They can, therefore, find good or optimal decompositions of reversible circuits with small width (4, perhaps 5 wires), but for more complex circuits with a large amount of wires, they have to give up. 
Heuristic methods using more compact function representations (\eg decision diagrams) have been suggested to make synthesis of larger functions possible~\cite{WilleDrechsler:2009,SoekenEtal:2010}. These, however, add many extra lines to the design, which are often left as garbage.

In parallel with the development in synthesis, methods for optimization of reversible circuits have been developed.
One such method is called \emph{template matching}~\cite{MillerMaslovDueck:2003,MaslovDueckMiller:2003}. Based on a large set of identity circuits, the method matches a subpart of the identity circuits with subparts of the circuit to be optimized. If a larger part of the identity circuit is found it can be replaced by the smaller part without changing the functionality of the circuit. 
Other methods are optimizations based on ESOP minimization~\cite{StergiouEtAl:2004,VanRentergemDeVos:2007} and reducing logic depth by expanding the logic width (adding more lines)~\cite{MaslovSaeedi:2011,MillerWilleDrechsler:2010}.

\section{Hardware Description Languages}
\label{sec:cad:designLang}
Domain specific languages (DSLs) to describe computing systems and circuits have been extensively studied, \eg resulting in DSLs such as VHDL, Verilog, and SystemC.
Taking inspiration from this, Wille \etal implemented the reversible design language SyReC~\cite{WilleDrechsler:2010:book,WilleOffermannDrechler:2010}.
The language builds on Janus and, therefore, it has the same properties of being sequential and imperative. This makes it less suitable for describing logic with an inherent concurrent structure. Therefore, SyReC is mainly used for creating the logic-level data structures used in synthesis. It has been used to implement different circuits from multiplication to a simple architecture~\cite{OffermannEtal:2010,WilleEtal:2011,OffermannWilleDrechler:2011}.
Also, a student project at DIKU investigated a simple imperative language to describe reversible logic circuits~\cite{Lehnfeld:2010:StudRep}.

Our work focuses on using functional languages to describe reversible logic. Using functional languages to describe logic dates back to the 1980s (see `non-survey' by Sheeran~\cite{Sheeran:2005}), but although research has continued, these languages are not widely used in industry.
This, however, did not stop us. So we have designed two languages with the hope that description of reversible logic combined with functional languages can be a success.

The first language is a point-free combinator-style language and it is designed to be close to the reversible logic gate-level~\cite{Thomsen:2012:IFL}\paperref{paper:IFL}. The language is inspired by Sheeran's $\mu$FP~\cite{Sheeran:1984} but it is also related to other languages and models.
A first example is Ruby~\cite{JonesSheeran:1990}\footnote{Here, we refer to the hardware description language and \emph{not} the later dynamic-typed object-oriented language by Matsumoto.} that, even thought it is made to describe conventional circuits, also has algebraic laws for inverse composition.
From reversible computing, a computation model designed by Green and Altenkirch~\cite{GreenAltenkirch:2008} to study the relation between reversible and irreversible computations, use some similar basic combinators and some of the algebraic laws that is also used in our work.
James and Sabry's $\Pi^O$ calculus~\cite{JamesSabry:2012:POPL,JamesSabry:2012:RC} is a point-free language with a similar type system with product and sum of wires.
Finally, Coecke and Duncan's $\mathit{ZX}$-calculus~\cite{CoeckeDuncan:2008,CoeckeDuncan:2011} is a graphical calculus used to simulate quantum computations and uses some of the compositions (plus some special quantum compositions) and rewriting with using algebraic laws.

The second language is a linear typed higher-level functional language~\cite{Thomsen:2012:FDL} \paperref{paper:FDL} with constructs such as conditionals and a let-in statements for local wire updates, which uses size-change termination to ensure termination of recursions.
The language has some similarities with the previously mentioned reversible functional languages but also conventional languages like Lava~\cite{BjesseEtal:1998:Lava}
and Park and Im's linear lambda calculus ($l\lambda$)~\cite{ParkIm:2011}.

The \textbf{Paper~\ref{paper:FDL}} also shows ideas for a design flow that can be used garbage-free translation to reversible circuits, by using on the combinator language~\cite{Thomsen:2012:IFL} as an intermediate language.

%%%%%%%%%%%%%%%%%%%%%%%%%%%%%%%%%%%%%%%%%%%%%%%%%%%%%%%%%%%%%%%%%%%%%%%%%%%%%%%%
%%% Implementations of RC
%%%%%%%%%%%%%%%%%%%%%%%%%%%%%%%%%%%%%%%%%%%%%%%%%%%%%%%%%%%%%%%%%%%%%%%%%%%%%%%%
\chapter{Realization of Reversible Circuits}
\label{sec:circuits}
In Landauer's seminal paper~\cite{Landauer:1961}, he sets out to identify the possible sources of errors (or heat generation) in a physical computer. He identified three, where one of these was the dissipation of heat due to irreversibility, which then became the major topic of the paper.
The other two sources he identified as (1) incomplete switching due to fast switching time and (2) decay of stored information. This is a bit simplified but gives the overall picture. Landauer also knew that these two sources, in the implementation technology of his time (as well as today's CMOS), is much higher than Landauer's limit. So using reversible logic CMOS gates alone will, however, not be sufficient.

\section{Adiabatic Switching and Charge Recovery}
\label{sec:circuits:adiabatac}
The dissipative source relating to incomplete switching is what \emph{adiabatic switching} tries to overcome. The basic concept is to achieve asymptotically zero energy loss when the switching time goes towards infinite (hence the name adiabatic).\footnote{It is a common misunderstanding (and a often used reason for rejection of the entire field) that reversible and adiabatic circuits will lead to CMOS circuits that consume no energy at all. This is not true (it is not practical to use infinite time for a computation) and there is no claim of this in the literature. It is, however, a well-established fact (which to some extent also governs today's chip design of multi-core processors) that there is a tradeoff between energy consumption and switching time~\cite{ZhirnovEtAl:2003}.}

The concept has been studied throughout the 1980's by Fredkin and Toffoli~\cite{FredkinToffoli:1978:DARPA}, Mead~\cite{MeadConway:1980}, Feynman~\cite{Feynman:1996:lectures}, and Seitz \etal~\cite{SeitzEtal:1985:TR}, but it was Athas, Koller, and Svensson~\cite{AthasSvensson:1994,KollerAthas:1992} that first used the term adiabatic and applied adiabatic switching successfully to reversible logic by reusing the signal energy (sometime also called \emph{charge-recovery}). Reusing the input signals to generate the outputs is the key concept in Fredkin and Toffoli's conservative logic; you have the same number of billiard balls at input and output. Reversible logic can easily be converted to conservative logic by using complementary signals: each bit is represented by both its value and its negated value.
The combination of adiabatic switching and charge recovery is the key concept behind the designed (and fabricated) logic families that followed. We will here sketch two different families (perhaps the two most influential families  to reversible logic, so far), but there exist others, \eg Kramer \etal~\cite{KramerEtal:1995}, Vetuli \etal~\cite{VetuliEtal:1996}, and Amirante \etal~\cite{AmiranteEtal:2002:fp}.

\emph{Split-level charge recovery} (SLCR) logic was presented in 1994 by Younis and Knight~\cite{YounisKnight:1994} with some later improvements by Frank~\cite{Frank:1999}. Inspired by work by Hall~\cite{Hall:1993}, the gates resemble those of static CMOS logic, but with the significant difference that the constant voltage-source and voltage-drain are exchanged with trapezoid-shaped clock-signals. Full reversibility can, therefore, \emph{not} be achieved at gate level, but instead Bennett's compute-copy-uncompute method is applied to ensure charge recovery. The Pendulum processor was implemented in this logic family~\cite{Vieri:1999}.

\emph{Complementary pass-transistor} (CPT) logic was developed by De Vos~\cite{DeVos:2010:book,DeVos:1999} a few years after SLCR and has some similarities with work by Seitz \etal~\cite{SeitzEtal:1985:TR} and Merkle~\cite{Merkle:1993:REL}. Here, the reversible gates are implemented by pass-gates, which work as switches that `guide' the dual-line input signals to the desired output lines. The adiabatic switching is achieved using signals that gradually change from `undefined' to either \true or \false and back (often using a triangular, trapezoidal, or sine-wave signal), and the pass-gates, thus, switch only when there is little voltage across the gate. As pass-gates are not \emph{ideal} switches, this logic family has also been called semi-adiabatic~\cite{Frank:2003}. An advantage of using CPT logic is, however, that the circuits can directly be used in both directions, showing reversibility directly.

SPICE simulations of reversible CPT circuits have shown that such implementations have the potential to reduce energy consumption by about a factor of ten~\cite{DeVosVanRentergem:2005} using 0.35 $\mu$m CMOS. Similar results (with measurements showing a factor of about 5) have been presented by Amirante \etal~\cite{AmiranteEtal:2003} for their adiabatic logic family in 0.13 $\mu$m CMOS. In both cases with a `clock frequency' of about 10 MHz.

Our contribution has been to take the CPL logic family and implement the reversible gates applying the \emph{standard cells methodology}~\cite{Thomsen:2012:TR}\paperref{paper:techreport} with the goal of using them in a future design flow.
In 2003, Frank had the same goal with a generalized version of SLCR logic~\cite{Frank:2003}, but no results of this work have ever been published. Blotti and Saletti~\cite{BlottiSaletti:2004} have also looked at semi-custom designs for \emph{positive feedback adiabatic logic} (PFA logic), the family first presented by Vetuli \etal~\cite{VetuliEtal:1996}.

\section{Embedding in Static CMOS}
We do not expect to see fully reversible systems commercially available in the near future.
From this perspective it is interesting to consider hybrid systems, where reversible CMOS circuits are embedded within static CMOS.

The first to look at this were Amirante \etal~\cite{AmiranteEtal:2003,FisherEtal:2003} for PFA logic. Though the gate designs of this family are similar to SLCR logic, a trapezoidal signal is used to switch the transistors. This result in the problem of converting between a trapezoidal and a digital signal; a problem we also have with CPT logic. Amirante \etal use a two-stage memory to synchronize the digital input with desired clock and two 2-to-1 switches to generate the non-inverted and inverted trapezoidal signals.

We have taken a similar approach~\cite{BurignatEtal:2012:LNCS}\paperref{paper:interface}, but a major difference is that we also want to be able to use the reversible circuits in both directions. We solved this by an extra array of parallel switches that is controlled by a direction bit. In the implemented design, all signals (inputs, direction bit, trapezoidal signals) is generated by an FPGA.

%%%%%%%%%%%%%%%%%%%%%%%%%%%%%%%%%%%%%%%%%%%%%%%%%%%%%%%%%%%%%%%%%%%%%%%%%%%%%%%%
%%% Future work
%%%%%%%%%%%%%%%%%%%%%%%%%%%%%%%%%%%%%%%%%%%%%%%%%%%%%%%%%%%%%%%%%%%%%%%%%%%%%%%%
\chapter{Conclusions and Perspectives}
\label{sec:conclusion}
In this thesis, we have investigated the feasibility of designing and implementing garbage-free reversible computing systems. We have found that this, to a large extent, is possible with the knowledge we have today, but there are still many non-trivial barriers that need to be overcome. Experience and `expert knowledge' about reversible computing is definitely an advantage when making these designs, but this is, of course, also the case with many other areas of computer science.

More specifically, we have developed new garbage-free circuits for addition and are working towards a general multiplication circuit. We have also combined multiple operations together to implement a reversible arithmetic logic unit.
With these and other garbage-free arithmetic circuits it is possible to design larger reversible computing systems. As an example, we have implemented discrete lossless transforms by redesigning these with a lifting scheme. We have also shown the design of a reversible computing architecture and implemented this using only reversible logic gates.
While, these are still small systems, with further development it should be possible to use similar strategies to implement even larger systems.

From our own design experience, we know that designing logic gate-level circuits quickly becomes complicated when the functionality and number of wires involved are increased.
To make the design process easier, we have developed two hardware description languages. Using  examples from known reversible circuits, we have shown that circuits can be described reasonably concisely. These are, however, still small examples and we need to implement a larger system to show the usefulness of the languages.

There are basically two different gains that are advocated for the use of reversible systems.
The first is reduced energy consumption both due to Landauer's principle and a change to a adiabatic CMOS logic family or another future technology.
The second gain is functionality due to the fact that the circuits (and programs) can be used in both directions. The advantage is that the same implementation can be used at multiple places (\eg the same design for both the fast fourier transform (FFT) and inverse FFT) or that the same physical circuit can be used for multiple purposes (\eg if the FFT is little used the same circuits can also be used for the inverse FFT).
While realizing the first gain is left for future development, we believe that, with our design experience, the second gain is possible to achieve today, with benefit to future reversible computing systems.

\section{Future Work}
\label{sec:conc:futureWork}
Though the foundations for reversible computing were laid fifty years ago, and interest have increased considerably in recent years, in many ways the area is still young.
In this PhD thesis, some of the unknown land was covered, but there is still plenty to explore for the future.

\subsection{Arithmetic Logic Circuits}
For conventional logic circuits there exist much research, even whole books, dedicated to the design and implementation of computer arithmetic. This is definitely not the case for reversible logic.
The constraint that the circuits must be garbage-free is what makes it an interesting research problem, but most proposed designs (both hand-made and CAD generated) still implement the conventional algorithms \emph{with} garbage. They use the reversible gates, but as their sole goal is to reduce logic size or number of garbage bits for a specific fixed-size circuit, very little knowledge is actually gained from this approach.

However, arithmetic functions often have some inherent properties that can be exploited to make a very regular circuit design. A good example is the ripple-carry adder, where only a redesign gave the garbage-free V-shaped adder; a redesign that none of the automatic approaches can find.
In many cases the arithmetic function itself must also be redefined, such that it can be expressed reversibly. Here our current work on multiplication is the obvious example.
With this in mind, we need more design work on good garbage-free implementations of reversible circuits.

\subsection{Computer Aided Design}
A lot of research has focused on reversible (or quantum) logic synthesis, resulting in optimal circuits for small input sizes.
There has, however, been very little research on how to describe and design reversible systems; often a truth table (a permutation of the input vectors) is used.

Recently, SyReC~\cite{WilleOffermannDrechler:2010} and my two languages~\cite{Thomsen:2012:FDL,Thomsen:2012:IFL} have been presented, but these are still only initial steps and far from `full' description languages. More work on these languages is needed to make them better to use for descriptions of reversible circuits. A good way to acquire experience is by implementing reversible systems in these languages.

Also, algorithms for optimization have only been designed for gate-level descriptions. We must move this to a higher abstraction level and possible methods for this could be term rewriting or partial evaluation. However, there exist many other conventional optimization methods (\eg from compiler technology) and some of these might also apply.

We know that there is a tradeoff between ancilla lines (logic width), logic depth, and the number of gates; \eg adding one ancilla line allows a linear-depth adders with few gates, while adding $n$ ancilla lines allows a logarithmic-depth adder. If we can find a more exact relations between these resources we can use this in synthesis. The descriptions in reversible HDLs have some degree of modularity, so we could also use approaches for trading logic depth with adding~\cite{MillerWilleDrechsler:2010,MaslovSaeedi:2011} or removing~\cite{WilleSoekenDrechsler:2010} ancilla lines, depending on the ancillae lines already available.

Feynman's widely used diagram notation has an inherent 1-dimensional structure, in the sense that gates only operate vertically with computations proceeding from left to right. This 1-dimensional structure is a good description of many quantum architectures, but in recent years new architectures have been suggested, which have a 2 or even 3-dimensional structure. This has led to research in quantum circuit that use more dimensions~\cite{ChoiVanMeter:2011:QID,Rosenbaum:2012:NNA,PhamSvore:2012:RC}. In quantum circuits design there is a `nearest-neighbor'-approach coming from quantum architecture models, in which qubits can only interact with its neighboring qubits.
For CMOS circuit there is no nearest-neighbor problem, but the circuit are 2-dimensional and methods for placing and routing these circuits have been used for many years. There is, however, an important difference between these quantum architectures and CMOS circuits: qubits are represented with a single object (in some quantum architectures the qubits are even fixed in a placed) and the operations interact between them, while in CMOS the gates (operations) are places and the wires (bits) are routed in-between. It would, however, still be interesting to see if part of the place-and-route methods can be applied to quantum circuits also.

\subsection{Implementation of Circuits}
Only very recently was Landauer's principle experimentally verified~\cite{BerutEtal:2012:Nat}, but it is still any open question how (or if) this can be used for energy reduction in a `real' computer.
Initial simulations of adiabatic switched CMOS circuits show a possible energy reduction~\cite{DeVosVanRentergem:2005,AmiranteEtal:2003}, but there is still no experimental evidence for a whole system.
Very recently Orlov \etal~\cite{OrlovEtal:2012} showed that reversibly modifying (with copy and uncopy operations) a simple memory element (a capacitor and a resistor) with adiabatic switching can be done with lower heat dissipation than Landauer's limit at up to about 15 MHz.
It is, however, a possibility that CMOS technology will never switch efficiently enough for implementations of reversible gates to be viable, if reduction of energy consumption is the main goal.
It could be that a completely new implementation technology is needed~\cite{WeslerEtal:2008}. Some potential technologies are nanoelectronic devices~\cite{GalatsisEtal:2009},
nanomagnets \cite{LambsonEtak:2011}, and superconductor electronics~\cite{Likharev:2012}.
This work I will, however, leave to qualified engineers and physicists.

\subsubsection{Asynchronous Circuits}
\emph{Asynchronous circuits} \cite{Sparso:2006:book,Lines:1998} have a long history as a technology that show promising results with respect to speed and energy consumption, but it is also a technology that, so far, have had little influence outside the research environments.\footnote{Asynchronous circuits have been used for commercially produced network switches, which have also lead to interest from Intel~\cite{Lines:2004}.} In this sense it has a comparable history to using functional hardware description languages, and to some part also adiabatic (reversible) circuits.
It would be very interesting to investigate if a combination of asynchronous and reversible (adiabatic) circuits would be a good match. Describing and synthesizing asynchronous circuits is hard (there exist some CSP-based approaches), so perhaps a combination with functional languages is even possible.

%%%%%%%%%%%%%%%%%%%%%%%%%%%%%%%%%%%%%%%%%%%%%%%%%%%%%%%%%%%%%%%%%%%%%%%%%%%%%%%%
%%% BIBLIOGRAPHY
%%%%%%%%%%%%%%%%%%%%%%%%%%%%%%%%%%%%%%%%%%%%%%%%%%%%%%%%%%%%%%%%%%%%%%%%%%%%%%%%

\cleardoublepage
\renewcommand{\sc}[1]{\textsc{#1}}
\bibliographystyle{acm}
\bibliography{bibliography}

% %%%%%%%%%%%%%%%%%%%%%%%%%%%%%%%%%%%%%%%%%%%%%%%%%%%%%%%%%%%%%%%%%%%%%%%%%%%%%%%%%%% RESEARCH PAPERS
% %%%%%%%%%%%%%%%%%%%%%%%%%%%%%%%%%%%%%%%%%%%%%%%%%%%%%%%%%%%%%%%%%%%%%%%%%%%%%%%%%\cleardoublepage
\appendix

\newcommand{\remove}[1]{}
\makeatletter
\newcommand\paper{\@startsection{section}{1}{\z@}%
                                  {\z@}%
                                  {2.3ex \@plus.2ex}%
                                  {\remove}}
\makeatother

\newcommand{\bref}[1]{\textbf{Paper~\ref{#1}:}}
\cleardoublepage
\chapter{Papers on Gate-Level Designs of Arithmetic Functions}
\label{papers:RevLogDes}
This appendix contains three papers relating to reversible logic design of arithmetic circuits.

\begin{description}
\item[\bref{paper:JSA}] Thomsen, M.K., Gl\"uck, R.: Optimized Reversible Binary-Coded Decimal Adders. In: Journal of Systems Architecture, vol. 54, issue 7, pp. 697--706, 2008. \copyright~ Elsevier B.V. 2008

\item[\bref{paper:PPL}] Thomsen, M.K., Axelsen, H.B.: Parallelization of Reversible Ripple-Carry Adders. In: Parallel Processing Letters 19(2), pp. 205--222, 2009. \copyright~ 2009 World Scientific Publishing Company. 

Preprint version.

\item[\bref{paper:RC2012:Mult}] Axelsen, H.B., Thomsen, M.K.: Garbage-Free Integer Multiplication with Constants. In: Gl\"uck, R., Yokoyama, T. (eds.) 4th Workshop on Reversible Computation (RC), Preliminary Proceedings, pp. 198--204, 2012.

\end{description}

\paper{Optimized Reversible Binary-Coded Decimal Adders}
\label{paper:JSA}

\paper{Parallelization of Reversible Ripple-Carry Adders}
\label{paper:PPL}

\paper{Garbage-Free Integer Multiplication with Constants}
\label{paper:RC2012:Mult}

\chapter{Papers on Reversible Architectures} \noindent
\label{papers:RevArch}
This appendix contains two papers relating to design of reversible architectures.

\begin{description}
\item[\bref{paper:JPhysA}] Thomsen, M.K., Gl\"uck, R., Axelsen, H.B.: Reversible arithmetic logic unit for quantum arithmetic. Journal of Physics A: Mathematical and Theoretical 43(38), 382002(10pp), 16 Aug 2010. \copyright~ 2010 IOP Publishing Ltd.

\item[\bref{paper:Bob}] Thomsen, M.K., Axelsen, H.B., Gl\"uck, R.: A reversible processor architecture and its reversible logic design. In: De Vos, A., Wille, R. (eds.) Reversible Computation, RC2011, Revised Papers. LNCS, vol. 7165, pp. 30--42, 2012. \copyright~ Springer-Verlag Berlin Heidelberg 2012.

\end{description}

\paper{Reversible Arithmetic Logic Unit for Quantum Arithmetic}
\label{paper:JPhysA}

\paper{A Reversible Processor Architecture and its Reversible Logic Design}
\label{paper:Bob}

\chapter{Papers on Implementation of Reversible Linear Transforms} \noindent
\label{papers:RevHighDes}
This appendix contains two papers relating to design of linear transforms in reversible logic.

\begin{description}
\item[\bref{paper:Multimedia1}] De Vos, A., Burignat, S., Thomsen, M.K.: Reversible Implementation of a Discrete Integer Linear Transform. In: Journal of Multiple-Valued Logic and Soft Computing, Special Issue: Reversible Computation, vol. 18, issue 1, pp. 25--35, 2012. \copyright~ Old City Publishing

Preprint version. This paper is the journal version of the paper from the 2nd Workshop on Reversible Computation, RC2010.

\item[\bref{paper:Multimedia2}]  Burignat, S., Vermeirsch, K., De Vos, A., Thomsen, M.K.: Garbageless Reversible Implementation of Integer Linear Transformations. In: Gl\"uck, R., Yokoyama, T. (eds.) 4th Workshop on Reversible Computation, Preliminary Proceedings, pp. 198--204, 2012.

\end{description}

\paper{Reversible Implementation of a Discrete Integer Linear Trans-form~~}
\label{paper:Multimedia1}

\paper{Garbageless Reversible Implementation of Integer Linear Transformations}
\label{paper:Multimedia2}

\chapter{Papers on Design Languages for Reversible Logic} \noindent
This appendix contains two papers on description languages for reversible logic.

\begin{description}
\item[\bref{paper:IFL}] Thomsen, M.K.: Describing and Optimising Reversible Logic using a Functional Language. In: Gill, A., Hage, J. (eds.) Implementation and Application of Functional Language, IFL2011. LNCS, 2012. \copyright~ Springer-Verlag Berlin Heidelberg 2012.

Preprint version.

\item[\bref{paper:FDL}] Thomsen, M.K.: A Functional Language for Describing Reversible Logic. In: Forum on Specification \& Design Languages, 2012

Close-to-preprint version.

\end{description}

\paper{Describing and Optimising Reversible Logic using a Functional Language}
\label{paper:IFL}

\paper{A Functional Language for Describing Reversible Logic}
\label{paper:FDL}

\chapter{Papers on Engineering of Reversible Circuits} \noindent
This appendix contains two papers on engineering of reversible circuits.

\begin{description}
\item[\bref{paper:techreport}] Thomsen, M.K.: Design of Reversible Logic Circuits using Standard Cells / Standard Cells and Functional Programming. DIKU technical report. No. 2012-03, ISSN 0107-8283, 2012.

\item[\bref{paper:interface}] Burignat, S., Thomsen, M.K., Klimczak, M., Olczak, M., De Vos, A.: Interfacing Reversible Pass-Transistor {CMOS} Chips with Conventional Restoring {CMOS} Circuits. In: De Vos, A., Wille, R. (eds.) Reversible Computation, RC2011, Revised Papers. LNCS, vol. 7165, pp. 112--122, 2012. \copyright~ Springer-Verlag Berlin Heidelberg 2012.

\end{description}

\paper{Design of Reversible Logic Circuits using Standard Cells / Standard Cells and Functional Programming}
\label{paper:techreport}

\paper{Interfacing Reversible Pass-Transistor CMOS Chips with Conventional Restoring CMOS Circuits}
\label{paper:interface}

\end{document}